\documentclass[journal=ancac3,manuscript=article]{achemso}

\usepackage[version=3]{mhchem} 
\usepackage{caption} 

\author{Dennis Palagin}
\email{dennis.palagin@ch.tum.de}
\author{Tobias Teufl}
\author{Karsten Reuter}
\affiliation[TUM]
{Department Chemie, Technische Universit{\"a}t M{\"u}nchen, Lichtenbergstr. 4, D-85747 Garching, Germany}

\title[Multi-Doping of Si Cages]
{Multi-Doping of Si Cages: High Spin States beyond the Single-Dopant Septet Limit}

\begin{document}

\begin{abstract}

Density-functional theory based global geometry optimization is employed to systematically scrutinize the possibility of multi-doping of hydrogenated Si clusters in order to achieve high spin states beyond the septet limit of a single-atom dopant. While our unbiased configurational search reveals that the previously suggested Si$_{18}$H$_{12}$ double hexagonal prism structure is generally too small to accommodate two dopants in magnetized state, the larger Si$_{24}$H$_{24}$ cage turns out to be suitable for such applications. For dimer dopants $M$$_{2}$$^{+}$ $=$ Cr$_{2}$$^{+}$, Mn$_{2}$$^{+}$ and CrMn$^{+}$, the structural integrity of the host cage is conserved in the ground-state structure of corresponding $M$$_{2}$$^{+}$@Si$_{24}$H$_{24}$ aggregates, as is the unusually high spin state of the guest dopant, which in case of Cr$_{2}$$^{+}$ already exceeds the single-atom dopant septet limit by almost a factor of two. Moreover, the possibility of further increasing the cluster spin moment by encapsulating an even larger number of dopants into a suitably sized hydrogenated Si cage is illustrated for the example of a (CrMn$^{+}$)$_{2}$@Si$_{28}$H$_{28}$ aggregate with a total number of 18 unpaired electrons. These results strongly suggest multi-doping of Si clusters as a viable route to novel cluster-based materials for magneto-optic applications.


\end{abstract}



\begin{figure*}
\includegraphics[width=12cm]{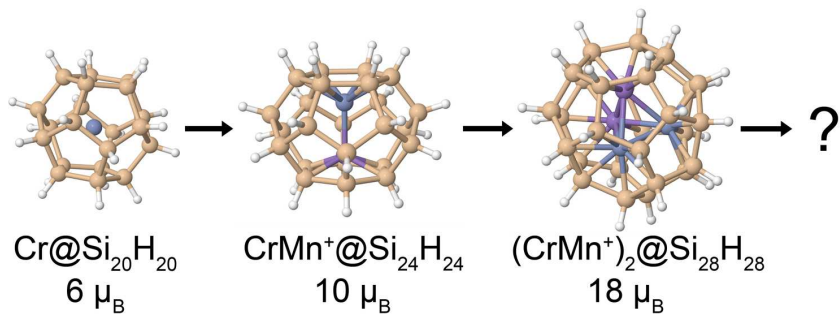}
\caption*{Table of Contents Graphic}
\label{TOC_graphic}
\end{figure*}

\section{Introduction}

Assembly at the molecular level is one of the most fundamental approaches in chemistry for obtaining new materials with desired novel properties, for instance relevant for the growing area of molecular electronics \cite{Reed_1997, Nakamura_2003, Imahori_2004, Deibel_2010}. For such nano-assembly, the use of endohedrally doped clusters as building blocks proved to be fruitful due to their suitable cage-like geometries, high stability and easily tunable electronic and optical properties \cite{Castleman_Khanna_2009, Claridge_2009}. Adding several different dopant atoms to each building block opens another configurational dimension and was already shown to yield unique properties unavailable for singly-doped clusters, such as large dipole \cite{Iwasa_Nakajima_2012} or magnetic \cite{Westerstrom_2012} moments. To date, two main directions of such research have been followed: either to form multiply doped systems by creating hetero-oligomers \cite{Iwasa_Nakajima_2012, Kumar_PRL_2001, Torres_JPCC_2011, Torres_Int_Q_2011, Balbas_PRB_2011, Palagin_2013} or even silicon nanorods \cite{Menon_2002, Kumar_2002, Andriotis_2002} made of several singly-doped clusters, or to create truly multiply doped individual cages, each accommodating several dopant atoms \cite{Westerstrom_2012, Sato_2012, Ge_2005, Yamada_2005} or even whole molecules \cite{Kurotobi_2011, Sabirov_2013} within the same cavity. Along the latter route, which is of much interest for magneto-optic applications, incorporation of magnetic ions within the cluster cage was demonstrated to lead to formation of so-called single-molecule magnets with long relaxation times \cite{Westerstrom_2012}.

To the best of our knowledge, such ideas have only been pursued for carbon fullerene cages, judged as being sufficiently large and stable to accommodate complex dopants. However, hydrogenated Si clusters are also capable of forming large fullerene-like cages \cite{kumar03, kumar_prb_2007, kumar_prb_2009, palagin12}, which would then offer comparable space to create multi-doped endohedral Si structures. Apart from the different material class, an advantage of hydrogenated silicon cages over carbon ones is hereby the prospect of bringing the unique properties of the encapsulated dopants to cluster-assembled materials \cite{Claridge_2009}, for example by constructing doubly Si--Si bound aggregates as recently suggested \cite{Palagin_2013}. 

An exciting prospect of multi-doping within the same cluster cavity is the potentially increased magnetic moment of such a complex dopant. While single-atom transition metal dopants are limited by the maximum number of five ($d$-shell) or six ($s$- and $d$- shells) unpaired electrons (e.g. 4$s$$^{2}$3$d$$^{5}$ or 4$s$$^{1}$3$d$$^{5}$ configurations of Mn and Cr, respectively), complex dopants open a path to larger spin moments. For instance, Lau \emph{et al.} reported experimental evidence for localized character of valence electrons in Cr$_{2}$$^{+}$, Mn$_{2}$$^{+}$ and CrMn$^{+}$ dimer cations \cite{Lau_2009}. With bonding predominantly mediated by $4s$ electrons, these transition-metal dimers already exhibit local high spin states with up to 5 $\mu_{B}$ spin moment of the $3d$ electrons at each metal core. If this is preserved upon encapsulating these dimers within a Si cage, such unique magnetic properties can be transported to new assembled materials.

In order to systematically assess the possibility of multi-doping of Si clusters, here we apply density-functional theory (DFT) based global geometry optimization to validate the idea of stabilizing symmetric endohedral structures with high magnetic moments. While $M$$_{2}$Si$_{18}$$^{+}$ has been experimentally identified as smallest Si structure capable of encapsulating transition metal dimers \cite{kumar03, Janssens_2007, Ji_Luo_2012}, we find that even at the minimized dopant-cage interaction of the hydrogenated $M$$_{2}$Si$_{18}$H$_{12}$$^{+}$ the spin moments of the appealing Cr$_{2}$$^{+}$, Mn$_{2}$$^{+}$ and CrMn$^{+}$ magnetic dimers are completely quenched. This is much different when going towards larger hydrogenated Si cages (Si$_{24}$H$_{24}$, Si$_{28}$H$_{28}$). Here, the identified ground-state structures indeed correspond to multi-doped endohedral cages with magnetic moments that go beyond the single-atom dopant 4$s$$^{1}$3$d$$^{5}$ septet limit.

\section{Computational Details}

All ground-state total energy calculations in this work have been performed with the all-electron full-potential DFT code FHI-aims\cite{blum09, ren12}. Electronic exchange and correlation was treated on the hybrid functional level with the PBE0 functional\cite{pbe0}. All sampling calculations were done with the ``tier 2'' atom-centered basis set using ``tight'' settings for numerical integrations. The stability of the identified minima has been confirmed by vibrational frequency analysis. For the subsequent electronic structure analysis of the optimized structures the electron density was recomputed with an enlarged ``tier 3'' basis set\cite{blum09}.

To ensure that the obtained geometries indeed represent the ground-state structures for all considered systems we relied on basin-hopping (BH) based global geometry optimization\cite{doye97, wales97_nature, wales00}, which samples the potential energy surface (PES) through consecutive jumps from one local minimum to another. In our implementation\cite{gehrke09, gramzow10, palagin11, palagin12} it is achieved by random displacement of atoms in the cluster in a so-called trial move followed by a local geometry optimization. A Metroplis-type acceptance rule is used to either accept or reject the jump into the PES minimum reached by the trial move.

Summarizing the computational details in the supplemental material, we here only point out that the employed DFT-based global geometry optimization approach has been rigorously evaluated in previous work on Si clusters \cite{Palagin_2013, palagin12, gehrke09, gramzow10, palagin11}, and has proven to both reliably predict geometries of lowest-lying isomers and to accurately describe their electronic structure. For the adequate discussion of the spin states manifold, calculations are carried out at the hybrid PBE0 \cite{pbe0} functional level, which is commonly agreed to yield rather reliable results for the rich transition metal chemistry \cite{Cramer_2009}. 

\section{Results and Discussion}

\subsection{\textit{M}$_{2}$Si$_{18}$H$_{12}$$^{+}$: the Smallest Multi-Doped Silicon Cluster}

Within this protocol, the starting point of our investigation is an extended configurational search, which confirms that the ground-state structure of Cr$_{2}$Si$_{18}$$^{+}$ is indeed a highly symmetric ($D_{6h}$) double hexagonal prism encapsulating the Cr$_{2}$$^{+}$ dimer, as suggested by Kumar \emph{et al.} \cite{kumar03} and Ji \emph{et al.} \cite{Ji_Luo_2012}, and in full agreement with the experimental results by Janssens \emph{et al.} \cite{Janssens_2007} However, the thorough electronic structure analysis revealed that the $sp^{3}$ Si dangling bonds are saturated through the strong Cr--Si interaction, which leads to the complete quenching of the Cr dopant high spin state \cite{supplement}. For instance, this can be clearly seen in the density of states (DOS) diagram, where the contribution of both Cr dopant atoms and the cage Si atoms to the frontier orbitals of the cluster is quite significant, which illustrates a strong interaction between the dopant and the cage. Quite similar DOS structure we have reported earlier for $M$Si$_{16}$$^{+}$ ($M$ $=$ Ti, V, Cr) cages \cite{palagin11}. In contrast, larger hydrogenated silicon cages, with the dopant metal isolated inside, exhibit negligibly small Cr--Si hybridization, which is illustrated by clear separation of the silicon and metal dopant levels in DOS \cite{palagin12, Palagin_2013}. This view is in line with the complete quenching of the Cr spin state within the Cr$_{2}$Si$_{18}$$^{+}$ structure and is also complemented by the analysis of the frontier orbitals shapes, where the contribution of Cr is quite pronounced. 

To check if the hydrogen termination idea to minimize the dopant-cage interaction \cite{kumar03, kumar_prb_2007, kumar_prb_2009, palagin12} is applicable in this case, we also ran a global geometry optimization of the corresponding fully hydrogen-terminated Cr$_{2}$Si$_{18}$H$_{12}$$^{+}$ structure. The obtained ground-state structure exhibits a heavily distorted geometry, while the symmetric and endohedrally doped hydrogenated double-prism is only a local minimum, energetically located 1.81 eV higher than the ground-state structure. This result agrees well with our previous suggestion of a generally insufficient space even within Si$_{16}$H$_{16}$ cages to conserve the magnetic moment of any $3d$ metal single atomic dopant \cite{palagin12}. While in case of Si$_{16}$H$_{16}$ the average radius of the cage is about 3 \AA, in the symmetric Cr$_{2}$Si$_{18}$$^{+}$ cluster the dopant-cage distance amounts only to 2.67 \AA, which renders the concept of conserving the atomic character of the magnetic dopant impossible. 

\subsection{Si$_{24}$H$_{24}$ Cage Doped with Magnetic Dimers}

\begin{figure}
\includegraphics[width=12cm]{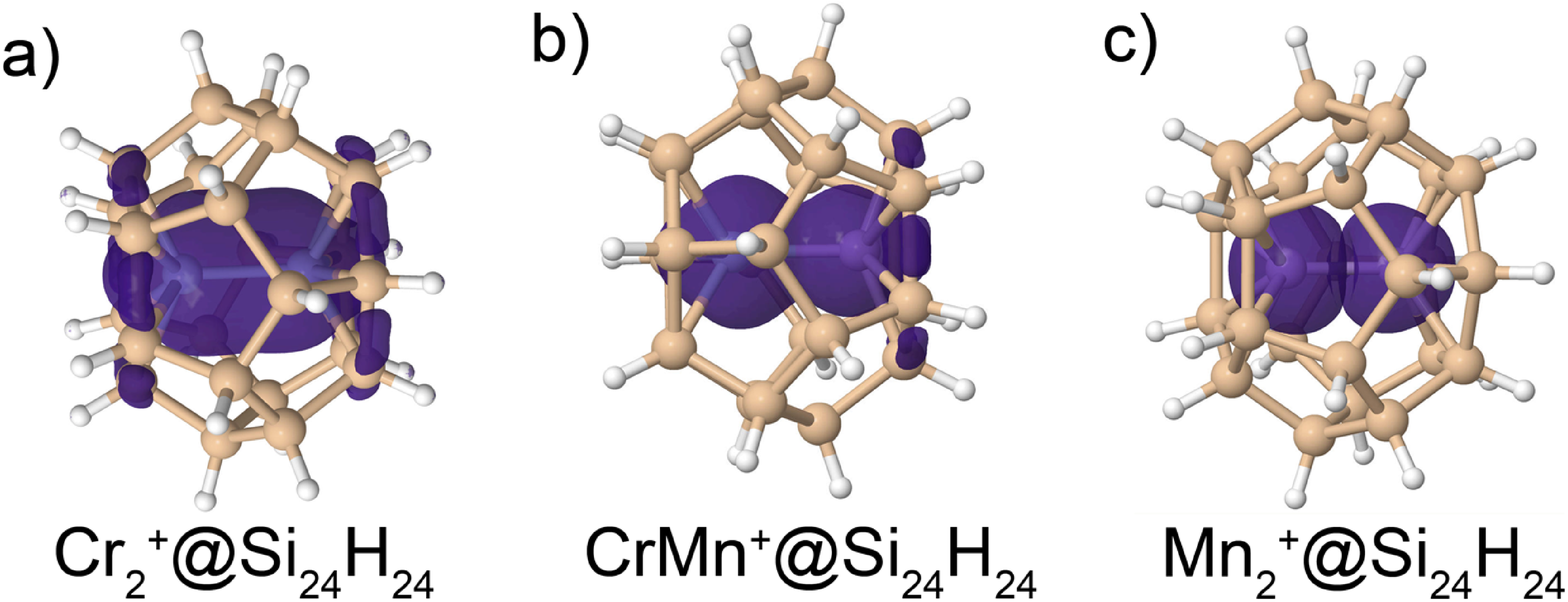}
\caption{Ground-state structures and spin density distribution (isosurfaces at 0.02  $e^{-}$/\AA$^{3}$) of a) Cr$_{2}$$^{+}$@Si$_{24}$H$_{24}$, b) CrMn$^{+}$@Si$_{24}$H$_{24}$, and c) Mn$_{2}$$^{+}$@Si$_{24}$H$_{24}$.}
\label{fig01}
\end{figure}

We thus proceed to take a look at the suitability of larger Si$_{n}$H$_{n}$ cages in stabilizing multiple magnetic dopants. As we have shown previously, Si$_{20}$H$_{20}$ is a smallest cage capable of conserving the atomic character of high-spin single-atom dopants such as Cr \cite{palagin12}. An intuitive candidate for multi-doping is then the next larger hydrogenated Si fullerene cage, i.e. Si$_{24}$H$_{24}$. As appealing dopant dimers we thereby stick to the Cr$_{2}$$^{+}$, Mn$_{2}$$^{+}$ and CrMn$^{+}$ series, for which a strong $3d$ valence electron localization and corresponding high-spin moments due to the half-filled $3d$ shells have been determined by core-level X-ray absorption spectroscopy experiments \cite{Janssens_2007} and are faithfully reproduced in our DFT calculations \cite{supplement}. These high-spin cation dimers thus seem to be perfect candidate dopants to yield highly magnetic building blocks for novel engineered materials. Global geometry optimization was correspondingly performed for $M$$_{2}$$^{+}$@Si$_{24}$H$_{24}$ clusters with $M$$_{2}$$^{+}$ $=$ Cr$_{2}$$^{+}$, Mn$_{2}$$^{+}$ and CrMn$^{+}$. In all cases the unbiased configurational search yields ground-state structures corresponding to highly symmetric endohedral $D_{6d}$ structures (Fig. \ref{fig01}) with a total of 11 (Cr$_{2}$$^{+}$), 10 (CrMn$^{+}$) and 9 (Mn$_{2}$$^{+}$) unpaired electrons. Interestingly, it is also possible to stabilize other spin states within the cage though. For example, a doublet Cr$_{2}$$^{+}$ dimer can also be encapsulated in CrSi$_{24}$H$_{24}$, oriented either along or perpendicular to the cage's principal axis \cite{supplement}. However, these structures possess much higher total energies (3.87 eV and 5.47 eV higher than the ground-state structure, respectively), which agrees well with the 4.24 eV difference between the corresponding high and low spin states of the isolated Cr$_{2}$$^{+}$ dimer. Another peculiarity is the remarkable agreement in the bond lengths of the isolated and encapsulated Cr$_{2}$$^{+}$ dimers. In the high-spin cage structure the Cr-Cr bond length is 2.83 \AA \, (compared to 2.90 \AA \, for isolated Cr$_{2}$$^{+}$ with 11 unpaired $e^{-}$), while for the two low-spin state structures the corresponding distances are 1.65 \AA \, and 1.60 \AA \, (compared to 1.56 \AA \, in the isolated doublet Cr$_{2}$$^{+}$ dimer).

\begin{figure}
\includegraphics[width=10cm]{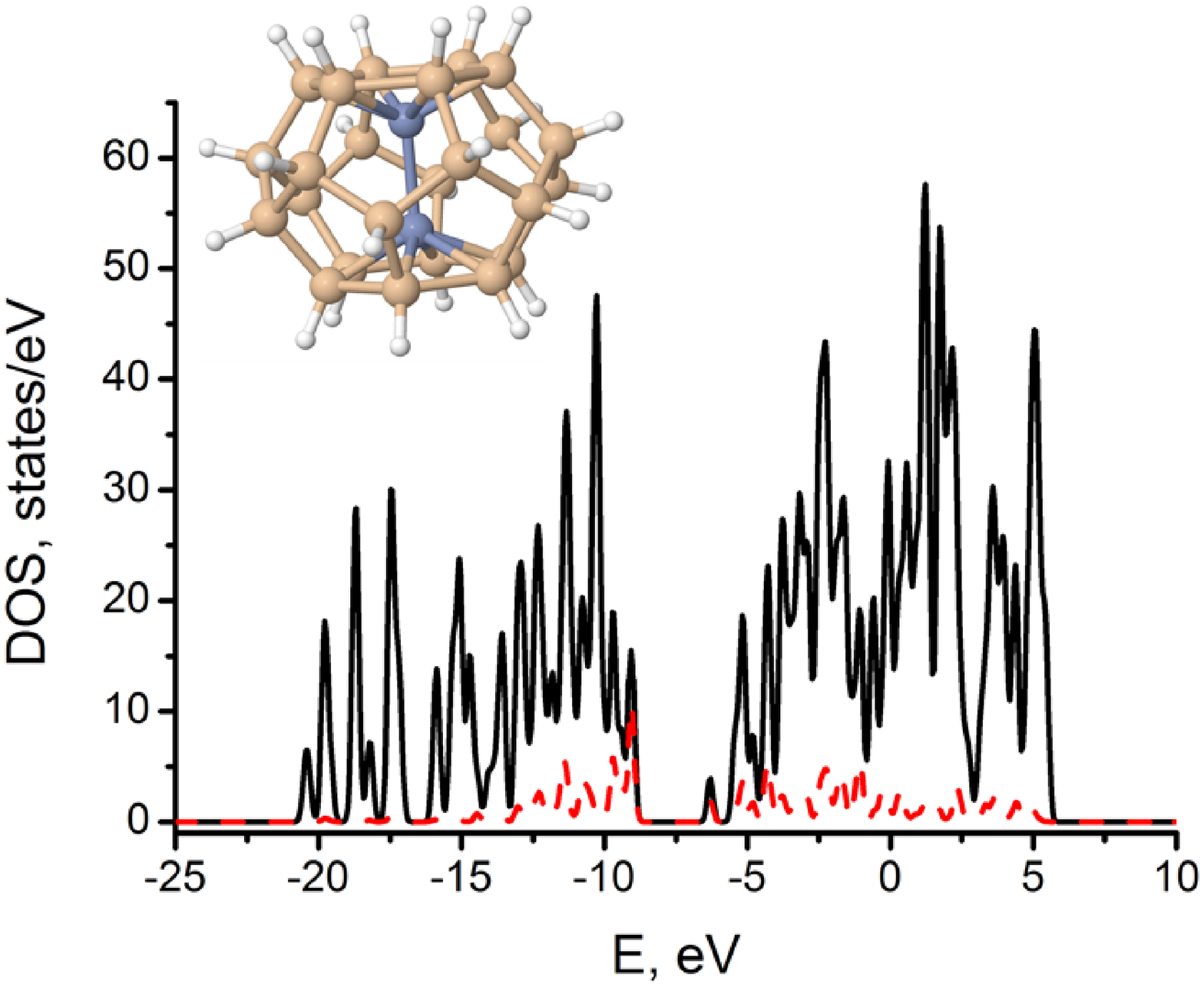}
\caption{Total density of states (black solid line) and DOS projected on the metal dopant (red dashed line) for Cr$_{2}$$^{+}$@Si$_{24}$H$_{24}$ (ground state) calculated at the hybrid PBE0 functional level. Highest occupied state lies at $-$9.04 eV, lowest unoccupied state lies at $-$6.30 eV; vacuum level is used as reference. The inset shows the identified ground-state geometry.}
\label{fig02}
\end{figure}

The stabilization of the high spin state of the encapsulated dimer in the Cr$_{2}$$^{+}$@Si$_{24}$H$_{24}$ ground-state structure, together with the reasonable agreement in both relative spin state energies and equilibrium distances between the encapsulated and isolated Cr$_{2}$$^{+}$ dimers, suggests that there should not be much interaction between the dopant dimer and the cage, i.e. the dopant character is more or less conserved within the cage. Notwithstanding, a rather high binding energy of 3.26 eV for the ground-state structure indicates the presence of sizable cage-dopant interaction that helps to stabilize the structure. In order to clarify this, we plot the spin density distribution within the Cr$_{2}$$^{+}$@Si$_{24}$H$_{24}$ ground-state structure (Fig. \ref{fig01}a) and analyze the density of states (DOS) diagram (Fig. \ref{fig02}). The spin density distribution in Fig. \ref{fig01}a clearly indicates that the unpaired electrons are predominantly located at the Cr atoms of the encapsulated dimer. However, there are also two notable protrusions of the spin density towards the two hexagonal facets of the cage located along the dimer axis. Some spin density is thus extended to adjacent Si--Si bonds through noticeable hybridization of the dimer frontier orbitals and Si cage states as is clearly discerned in the calculated system DOS (Fig. \ref{fig02}). 

\begin{figure}
\includegraphics[width=12cm]{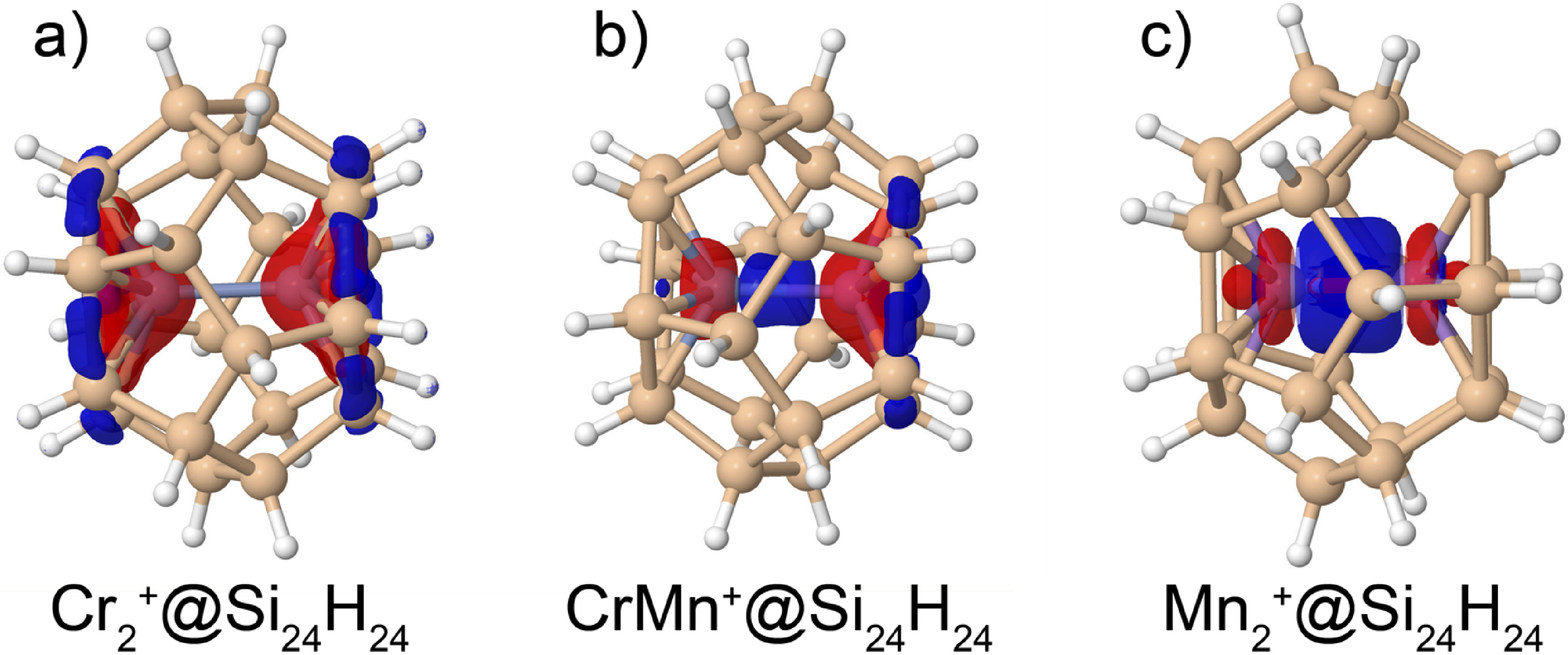}
\caption{Spin density difference for the identified ground-state structures of a) Cr$_{2}$$^{+}$@Si$_{24}$H$_{24}$, b) CrMn$^{+}$@Si$_{24}$H$_{24}$, and c) Mn$_{2}$$^{+}$@Si$_{24}$H$_{24}$.}
\label{fig03}
\end{figure}

This spin-density delocalization from dimer to cage is best visualized in a spin-density difference plot, where the spin density of the Cr$_{2}$$^{+}$ dimer and cage fragments are subtracted from the spin density of the Cr$_{2}$$^{+}$@Si$_{24}$H$_{24}$. The results in Fig. \ref{fig03}a clearly show that in the Cr$_{2}$$^{+}$@Si$_{24}$H$_{24}$ aggregate spin density is accumulated at the Si--Si bonds within the two hexagonal facets and reduced at the Cr cores as compared to an isolated Cr$_{2}$$^{+}$ dimer. A fraction of $d$-electrons located at the Cr cores in the isolated dimer is thus partially re-distributed to the area between the outer sides of the dimer and the hexagonal facets of the cage. Intriguingly, this does not affect the high spin state of the dopant, but apparently stabilizes the resulting aggregate, as also reflected by its large HOMO-LUMO gap value of 2.74 eV.

Figures \ref{fig01}b and \ref{fig01}c (as well as Figs. \ref{fig03}b and \ref{fig03}c) illustrate the identified ground-state structures and corresponding spin-density and spin-density difference distributions for the other two dopants (CrMn$^{+}$, Mn$_{2}$$^{+}$). Similar to the Cr$_{2}$$^{+}$ case the encapsulated CrMn$^{+}$ dimer has an equilibrium distance close to the one observed for the isolated dimer (2.80 \AA \, in both cases) and exhibits a high spin state of ten unpaired electrons. The analysis of its spin density distribution (Figs. \ref{fig01}b and \ref{fig03}b) shows that the ten unpaired electrons are again predominantly located at the metal cores with some additional spin-density delocalization towards the cage. Yet, this delocalization is now restricted to the Mn end of the heterodimer. This suggests a lowered dopant-cage interaction, which is indeed indicated by the computed smaller value of 2.67 eV for the binding energy and by the sharper metal-projected DOS peaks \cite{supplement}. This trend is continued for the Mn$_{2}$$^{+}$@Si$_{24}$H$_{24}$ aggregate, for which at a further reduced binding energy of 1.77 eV there is now no spin-density delocalization towards the Si cage at either end of the dimer (Figs. \ref{fig01}c and \ref{fig03}c). Through the resulting confinement the cage potential nevertheless does affect the internal electronic structure of the dimer, which in its encapsulated state exhibits a reduced number of only nine unpaired electrons. Similar to the situation for the isolated dimer, the additional pairing in such lower-spin states enforces the dimer bond and goes hand in hand with a corresponding contraction of the bond length; in the present case from 2.93 \AA \, in the isolated high-spin Mn$_{2}$$^{+}$ dimer to 2.59 \AA \, in the Mn$_{2}$$^{+}$@Si$_{24}$H$_{24}$ aggregate.

Overall these results thus demonstrate that already a Si$_{24}$H$_{24}$ cage is generally large enough to accommodate high-spin dimer dopants such as Cr$_{2}$$^{+}$, Mn$_{2}$$^{+}$ and CrMn$^{+}$. Intriguingly, the non-trivial cage-dopant interaction, still present in this smallest fullerene, does not compromise the symmetric cage structure in the ground-state geometry, but instead contributes to its stabilization. Most interesting from the perspective of conserving the character of the encapsulated dimer is thereby the CrMn$^{+}$@Si$_{24}$H$_{24}$ aggregate, where in contrast to Mn$_{2}$$^{+}$@Si$_{24}$H$_{24}$ the original high-spin state is maintained at a, compared to Cr$_{2}$$^{+}$@Si$_{24}$H$_{24}$, reduced cage-dopant interaction. We thus use this particular dimer to consider the next obvious question, namely whether it is possible to further increase the total number of unpaired electrons in the system by encapsulating several of these highly magnetic dimers into an even larger cage.

\subsection{Si$_{28}$H$_{28}$ Cage: Enough Space For Two?}

\begin{figure}
\includegraphics[width=10cm]{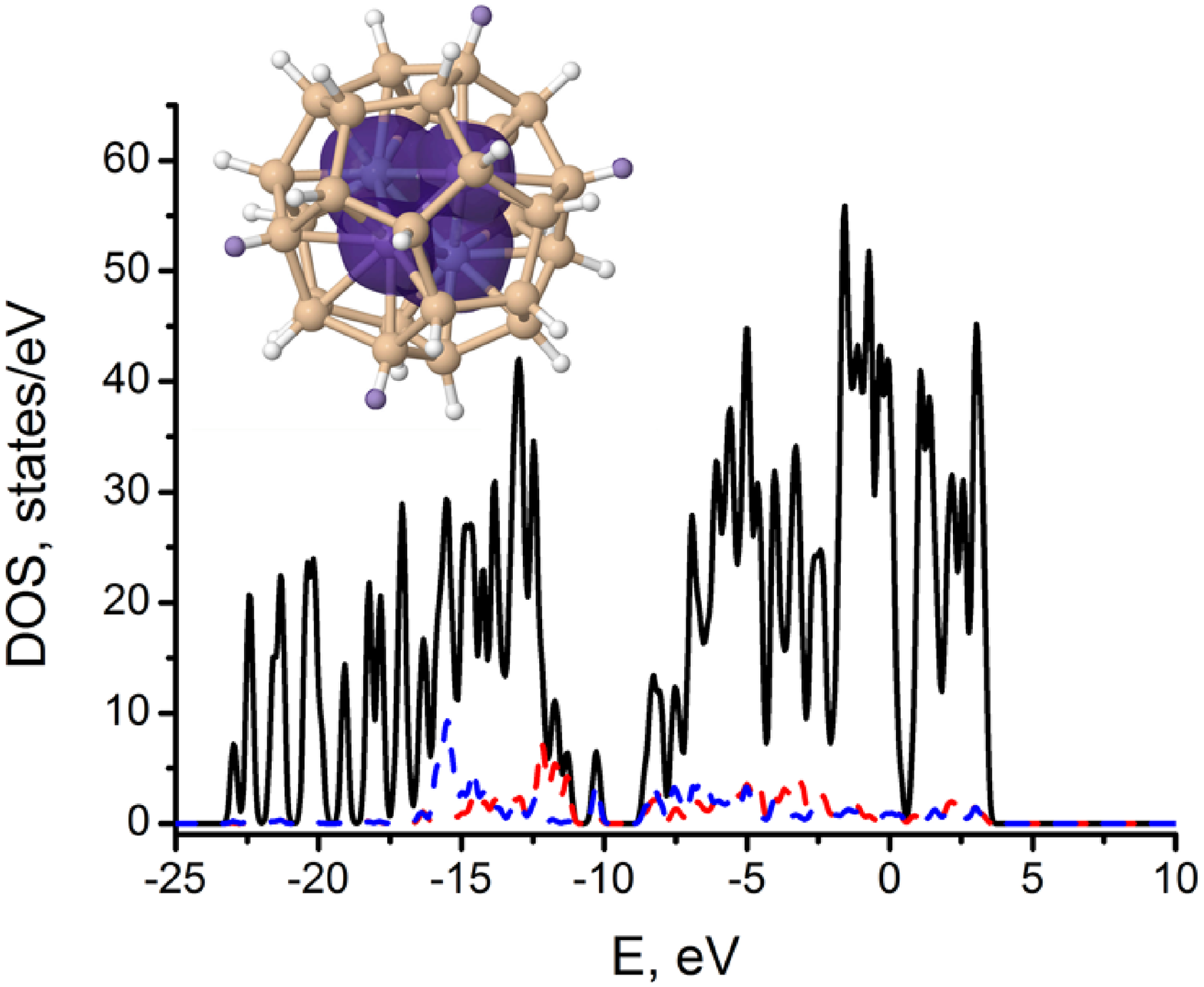}
\caption{Total (black), Cr-projected (red), and Mn-projected (blue) density-of-states for the (CrMn$^{+}$)$_{2}$@Si$_{28}$H$_{28}$ ground-state structure. All energies are given with respect to the vacuum level, with the HOMO located at $-$10.22 eV and the LUMO at $-$8.59 eV. The inset depicts the spin density distribution within the identified ground-state structure.}
\label{fig04}
\end{figure}

Numerically most demanding global geometry optimization was correspondingly run for a Si$_{28}$H$_{28}$ cage accommodating two CrMn$^{+}$ dopants. Indeed, our unbiased configurational sampling reveals that the (CrMn$^{+}$)$_{2}$@Si$_{28}$H$_{28}$ ground-state structure is a symmetrical $C_{2v}$ cage (see inset in the Fig. \ref{fig04}) with a total number of 18 unpaired electrons, energetically located 4.76 eV lower than the next distorted isomer. The spin-density distribution and DOS also shown in Fig. \ref{fig04} confirms that again all unpaired electrons are predominantly located at the four metal cores. 

A closer look at the DOS diagram (Fig. \ref{fig04}) reveals two distinct peaks corresponding to Cr-located and Mn-located unpaired electrons, confirming their localized character. Additional Mn peak at $-$10 eV shows partial redistribution of the Cr electrons towards Mn cores. No spin-density delocalization towards the cage is observed, which is consistent with the rather modest binding energy value of 1.71 eV. The thereby indicated low cage-dopant interaction suggests that the slightly lower spin moment of 18 unpaired electrons compared to the ideal 20 unpaired electrons of two isolated CrMn$^{+}$ dimers rather results from some intra-dopant charge rearrangement. In fact, the tetrahedral configuration adopted by the four metal cores in the (CrMn$^{+}$)$_{2}$@Si$_{28}$H$_{28}$ aggregate does in any case not justify a discussion in terms of two adjacent CrMn$^{+}$ dimers, with e.g. the Mn--Mn distance of 2.62 \AA \, being very similar to the 2.59 \AA \, found before in the Mn$_{2}$$^{+}$@Si$_{24}$H$_{24}$ case.

\section{Conclusions}

In summary, we have systematically assessed the possibility of multi-doping of hydrogenated Si clusters to obtain stable cage-like building blocks with high spin moments. Our unbiased first-principles global geometry optimizations show that it is possible to conserve both the structural integrity of the host cage and the high spin state of guest dimer dopants $M$$_{2}$$^{+}$ $=$ Cr$_{2}$$^{+}$, Mn$_{2}$$^{+}$ and CrMn$^{+}$ already for a Si$_{24}$H$_{24}$ fullerene. In the Cr$_{2}$$^{+}$@Si$_{24}$H$_{24}$ and CrMn$^{+}$@Si$_{24}$H$_{24}$ ground-state configuration the very high number of 11 and 10 unpaired electrons, respectively, exhibited by the isolated dimer is fully conserved and therewith exceeds the maximum septet limit achievable with a single-atom dopant almost by a factor of two. The crucial influence of the silicon cage size is thereby illustrated by the distorted ground-state structure obtained for the doubly-doped hydrogenated silicon cluster Cr$_{2}$Si$_{18}$H$_{12}$$^{+}$ previously suggested as smallest cage size capable of encapsulating metal dimers \cite{kumar03}. On the other hand, choosing sufficiently large cages allows accommodating even larger aggregates, as demonstrated for the combination of two CrMn$^{+}$ dimers inside a Si$_{28}$H$_{28}$ fullerene. The corresponding (CrMn$^{+}$)$_{2}$@Si$_{28}$H$_{28}$ ground-state structure exhibits a total number of 18 unpaired electrons, which is the highest magnetic moment of an endohedral cage reported so far.

This gives confirmation for the exciting perspective of tuning the magnetic moment by encapsulating a range of different dopants (starting from single atoms with different spin states, through highly magnetic dimers, and eventually up to more complex dopant aggregates) into a suitably sized hydrogenated silicon cage. Considering the possibility of obtaining larger fullerene-like Si cages \cite{johnston_book}, this suggests a path to tune the size of the hydrogenated Si cage to eventually accommodate dopants of virtually any size and complexity. Such highly magnetic doped clusters can then be used as building blocks for cluster-assembled materials, for example by constructing doubly Si--Si bound aggregates as we have recently suggested \cite{Palagin_2013}. Endohedral multi-doping of hydrogenated silicon fullerenes thus appears as a viable route to novel cluster-based materials for magneto-optic applications.


\acknowledgement

Funding within the DFG Research Unit FOR1282 and support of the TUM Faculty Graduate Center Chemistry is gratefully acknowledged.

\begin{suppinfo}

Detailed information on computational settings, as well as all details on the geometries and electronic structure analyses of the $M$$_{2}$Si$_{18}$$^{+}$ and $M$$_{2}$Si$_{18}$H$_{12}$$^{+}$ clusters, discussion of the preferable spin states of Cr$_{2}$, Cr$_{2}$$^{+}$, Mn$_{2}$, Mn$_{2}$$^{+}$ and CrMn$^{+}$ dimers, and the detailed analysis of the obtained geometries and stabilization mechanisms within $M$$_{2}$$^{+}$@Si$_{24}$H$_{24}$ clusteres can be found in the Supporting Information.

\end{suppinfo}


\bibliography{palagin_multidoping}

\end{document}